\documentclass[useAMS,fleqn,usenatbib]{mnras}

\usepackage{newtxtext,newtxmath}
\usepackage[T1]{fontenc}

\DeclareRobustCommand{\VAN}[3]{#2}
\let\VANthebibliography\thebibliography
\def\thebibliography{\DeclareRobustCommand{\VAN}[3]{##3}\VANthebibliography}
\usepackage{graphicx}	% Including figure files
\usepackage{amsmath}	% Advanced maths commands
\usepackage[utf8]{inputenc}
\usepackage{changepage}
\usepackage{afterpage}
\usepackage{floatpag}
\usepackage{transparent}
\usepackage{placeins}
\usepackage{physics}
\usepackage{booktabs}	
\usepackage{amsmath}	
\usepackage{ulem}
\usepackage{cancel}
\usepackage{xcolor}
\usepackage{grffile}
\usepackage{url}
\usepackage{comment}
\usepackage{natbib}
\usepackage{enumerate}
\usepackage{mathtools}
\usepackage{float}
\usepackage{soul}

\newcommand\empt{{}}

\newcommand\sect{Section}               
             
\newcommand\tabl{Table}                 
               
\newcommand\eqn{Eq.~}

\newcommand\fig{Fig.~}

\newcommand\msun{\mathrm{M}_{\odot}}
\newcommand\wlmass{M_{\rm{WL}}}
\newcommand\truemass{M_{\rm{true}}}

\newcommand\mfh{M_{500}}
\newcommand\mfhg{M_{500,\mathrm{G}}}
\newcommand\mfhm{M_{500,\mathrm{CoM}}}

\newcommand\rfh{r_{500}}
\newcommand\rmin{r_{\rm{min}}}
\newcommand\rmax{r_{\rm{max}}}

\newcommand{\gcentr}{\mathbf{\Theta}_{\mathrm{G}}}   % bottom om grav. pot. 
\newcommand{\mcentr}{\mathbf{\Theta}_{\mathrm{CoM}}} % center of mass
\newcommand{\szcentr}{\mathbf{\Theta}_{\mathrm{SZE}}} % SZ
\newcommand\bias{b}
\newcommand\logbias{\log{\bias}}
\newcommand\munormal{\mu_{\bias}}
\newcommand\sigmanormal{\sigma_{\bias}}
\newcommand\mulognormal{\mu_{\logbias}}
\newcommand\sigmalognormal{\sigma_{\logbias}}
\newcommand\overcorrection{\tau}

\newcommand\miscgmr{|\mcentr - \gcentr|_{\mathrm{(r)}}}

\newcommand\miscgsr{|\szcentr - \gcentr|_{\mathrm{(r)}}}

\newcommand\miscmsr{|\szcentr - \mcentr|_{\mathrm{(r)}}}

\newcommand\zlens{z_{\mathrm{l}}}
\newcommand\zsource{z_{\mathrm{s}}}
\newcommand\esource{\epsilon_{\mathrm{s}}}

\newcommand{\appropto}{\mathrel{\vcenter{
  \offinterlineskip\halign{\hfil$##$\cr
    \propto\cr\noalign{\kern2pt}\sim\cr\noalign{\kern-2pt}}}}}

\defcitealias{2024MNRAS.532.3359S}{S+24}

\defcitealias{2016A&A...594A..24P}{Planck Collaboration 2016}
\title[Non-isotropic mass bias]{Directional miscentering dependence in weak lensing mass bias}

\author[M. Sommer et al.]{Martin W. Sommer$^{1}$\thanks{E-mail: mnord@astro.uni-bonn.de (MWS)
},
Tim Schrabback$^{1,2}$,
and Sebastian Grandis$^{2}$,
\\
$^{1}$Argelander-Institut f\"{u}r Astronomie, Auf dem H\"ugel 71, D-53121 Bonn, Germany \\
$^{2}$Universität Innsbruck,  Institut für Astro- und Teilchenphysik,
Technikerstr. 25/8, 6020 Innsbruck, Austria \\
}

\date{(MNRAS accepted)}

\pubyear{\the\year{}}

\begin{document}
\label{firstpage}
\pagerange{\pageref{firstpage}--\pageref{lastpage}}
\maketitle

% Abstract of the paper
\begin{abstract}
Galaxy cluster masses estimated from parametric modeling of weak lensing shear observations are known to be biased by inaccuracies in observationally determined centers. It has recently been shown that such systematic effects can be non-isotropic when centers are derived from X-ray or Compton-Y (Sunyaev-Zeldovich effect) observations, which is often the case in practice. This fact challenges current methods of accurately correcting for weak lensing mass biases using simulations paired with isotropic empirical miscentering distributions, in particular as the effect on determined masses is currently a dominant source of systematic uncertainty. 
We use hydrodynamical cosmological simulations taken from the Magneticum Pathfinder simulations 
to show that the non-isotropic component of the mass bias can be reduced to within one percent of the mass when considering the center of mass, rather than the bottom of the gravitational potential, as the reference center of a galaxy cluster. 
%%% aims, methods, main results
%%% max 250 words (200 for letters)
%%%
\end{abstract}

% Select between one and six entries from the list of approved keywords.
% Don't make up new ones.

%%%%%%%%%%%%%%%%%%%%%%%%%%%%%%%%%%%%%%%%%%%%%%%%%%

\section{Introduction}
\label{sect:intr}

The distribution of clusters of galaxies, or halos, throughout the Universe traces the evolution of the matter distribution through space and time. In particular, the number density of galaxy clusters as a function of redshift and mass –- what is called the halo mass function -- depends critically upon the curvature of the Universe and the growth rate of primordial density fluctuations (e.g.~\citealt{2001ApJ...553..545H}), and thereby provide ways to investigate numerous aspects of the standard model of cosmology (e.g. \citealt{2011ARA&A..49..409A,2024arXiv240102075B,2024arXiv240208458G}). As masses are often inferred from mass-observable relations, the latter must in turn be calibrated through more direct means of measuring mass. 

The mass of a galaxy cluster is not uniquely defined, in part because the concept of a halo in itself has no strict definition. In the context of cosmology, a mass definition that can be exactly replicated in models and simulations is desirable. For this reason, the cluster mass is typically defined as the total mass inside a spherical region, defined such that the mean matter density inside this region is some overdensity factor $\Delta$ higher than either the critical or the mean mass density of the universe at the redshift of the halo. A mass $M_{\Delta}$ determined in this way then corresponds directly to a radius $r_{\Delta}$, given a suitably chosen center of the cluster. The choice of definition for said center coordinate, the accuracy with which it can be determined in observations, and systematic biases occurring from the lack of such accuracy are the central topics of investigation in this work. 

In modeling the halo mass function, a long-standing practice has been to use the bottom of the gravitational potential well of the halo as the center coordinate, in particular as this choice conforms to models of spherical infall \citep{1974ApJ...187..425P,1991ApJ...379..440B}. While such models were initially shown to be remarkably consistent with early n-body simulations \citep{2001MNRAS.321..372J}, more accurate models were later developed based on fits to simulations over large ranges of redshifts \citep[e.g.][]{1999MNRAS.308..119S,2005Natur.435..629S}. In more recent years, a process called mass function emulation, by which numerical simulations with varying parameters of the cosmological model are sampled, has been developed and refined \citep[e.g.][]{2006ApJ...646L...1H,2020ApJ...901....5B}. Given that the computation of the mass function at present is almost exclusively based on simulations, the definition of the halo center can be modified, as long as it can be uniquely defined from the mass distribution of a halo.

Because all mass acts gravitationally, arguably the most direct way of determining the mass of a halo observationally is through the distortion of background sources (galaxies) by its gravitational potential – the effect known as weak gravitational lensing (henceforth WL). There is a multitude of studies calibrating mass-observable relations in this way (e.g. \citealt{2014MNRAS.439...48A, 2014MNRAS.440.2077M}; \citetalias{2016A&A...594A..24P}; \citealt{2018MNRAS.474.2635S, 2019MNRAS.483.2871D,2019ApJ...878...55B,2019MNRAS.482.1352M,2021MNRAS.505.3923S,2022A&A...668A..18Z,2024A&A...687A.178G}). WL masses, in turn, suffer from a variety of systematic biases that must be corrected for, a problem that is receiving increasing attention as statistical uncertainties continue to decrease (e.g. \citealt{2019MNRAS.488.2041G}).

In this work, we are concerned with systematics pertaining to the accuracy of the weak lensing mass modeling, that is, we assume the data themselves are free of bias. We also neglect the effects of projection of uncorrelated matter along the line of sight \citep{2011MNRAS.412.2095H} and triaxiality (e.g. \citealt{2016MNRAS.457.1522A,2019MNRAS.483.2871D}), and the impact of parametric models of mass density (although we do use such a model). Our concern is purely with miscentering, that is, the impact of the choice of coordinates around which a specific azimuthally symmetric density model is taken. 

Significant efforts have been made towards quantifying the distribution of weak lensing mass bias distributions from simulations \citep{2011ApJ...740...25B,2011MNRAS.414.1851O,2012MNRAS.421.1073B,2017MNRAS.465.3361H,2018MNRAS.479..890L}, mostly considering cases in which no miscentering is present. More recent studies, accounting for miscentering effects (e.g. \citealt{2019MNRAS.483.2871D,2021MNRAS.505.3923S,2021MNRAS.507.5671G,2022A&A...668A..18Z,2023MNRAS.tmp..949C}), report estimates of residual systematic uncertainties on mass modeling bias between one and several percent, albeit without explicitly accounting for the possibility an anisotropic miscentering distribution.  \citet[henceforth~\citetalias{2024MNRAS.532.3359S}]{2024MNRAS.532.3359S} recently found that such effects may induce systematic errors of several percent using center coordinates derived 
either from peaks in the Compton-Y distortion (Sunyaev-Zeldovich effect, also SZE, \citealt{1970CoASP...2...66S,1980ARA&A..18..537S}) or centroids in the X-ray brightness -- both fairly standard approaches in practice \citep[e.g.][]{2021MNRAS.505.3923S,2023arXiv231012213B,2024arXiv240208456K}.

In this work, we shift the focus away from the observed cluster center, and towards the reference position; that is, the definition of the halo center. In particular, we explore the possibility of taking the center of mass, rather than the bottom of the gravitational potential, as the reference. Through simulations, we investigate whether non-isotropic effects in the weak lensing mass bias
can be reduced to a level below or similar to the expected statistical mass uncertainties of ongoing and planned large scale surveys. 

In \sect~\ref{sect:meth} we cover the necessary definitions for weak lensing observables and describe how lensing and SZE images are extracted from the simulated sample. The processes of generating and randomizing miscentering distributions for the study of anisotropies in the weak lensing mass bias are described, as well as the determination of halo mass given various center coordinates. We report our results in \sect~\ref{sect:resu}, discuss their implications and limitations in \sect~\ref{sect:disc}, and offer our conclusions in \sect~\ref{sect:conc}. 

We use the flat $\Lambda$CDM cosmological model of \citet{2011ApJS..192...18K}, with Hubble parameter $h = 0.704$ and total mass density $\Omega_{\rm{m}} = 0.272$.
Mass is quantified in terms of $\Delta = 500$, where $\Delta$ is the overdensity with respect to the critical density of the universe at the redshift of a galaxy cluster. 
We define the weak lensing mass bias as the ratio $\bias = \frac{\wlmass}{\truemass}$,
where $\wlmass$ and $\truemass$ are the measured and true masses, respectively. 
The natural logarithm is denoted $\log()$. 

\section{Method}
\label{sect:meth}

\subsection{Shear and convergence}
\label{sect:meth:wl}

The distortion of a background source at redshift $\zsource$ by a foreground lens (here, a cluster of galaxies) at redshift $\zlens$ is described in terms of the convergence $\kappa$ and the shear $\gamma$. The former is the surface mass density $\Sigma(\btheta)$ in units of the critical density
\begin{equation}
\Sigma_{\text{crit}} = \frac{c^2}{4 \pi G} \frac{1}{D_{\mathrm{l}}\beta}
\end{equation}
at projected position $\btheta$, where $c$ is the speed of light, $G$ is the gravitational constant and the lensing efficiency $\beta$ is defined as
\begin{equation}
\beta = \frac{D_{\text{ls}}}{D_{\text{s}}} H(\zsource-\zlens),
\end{equation}
with $D_{\mathrm{s}}$, $D_{\mathrm{l}}$, $D_{\mathrm{ls}}$ the angular diameter distances between observer and source, observer and lens, and lens and the source, respectively.\footnote{In the following, we shall take the positional argument $\btheta$ as implicit. }
The Heaviside step function, $H(\zsource-\zlens)$, is equal to one when $\zsource > \zlens$, and zero otherwise (reflecting the fact that lensing occurs only if the source is behind the lens). 

Shape distortions can be characterized by the reduced shear $g=g_1+\text{i}g_2$
through $g = \frac{\gamma}{1-\kappa}$,
where $\gamma$ is the (unobservable) complex shear $\gamma = \gamma_1+\rm{i}\gamma_2$ (for an in-depth account we refer to \citealt{2015RPPh...78h6901K}). For $|g| \leq 1$, the reduced shear can be estimated from averaged observed ellipticities\footnote{We define ellipticity as $\epsilon = (a-b)/(a+b) \cdot \mathrm{e}^{\mathrm{2i\phi}}$ for elliptical isophotes with minor-to-major axis ratio $b/a$ and
position angle $\phi$.} $ \epsilon = \epsilon_1 + \mathrm{i} \epsilon_2$, as \citep{1997A&A...318..687S}
\begin{equation}
    \epsilon = \frac{\esource + g}{1 + g^*\esource},
\end{equation}
where $g^*$ is the complex conjugate of the reduced shear, and $\esource$ is the intrinsic complex ellipticity of a source. Under the reasonable assumption that there are no preferred orientations among the sources, the expectation value of $\esource$ vanishes, so that the ellipticity is an unbiased estimator of the reduced shear. 

Shear, reduced shear and ellipticity can be decomposed into tangential (subscript $t$) and cross (subscript $\mathsf{x}$) components through 
\begin{equation}
\label{eq:shearxt}
\begin{array}{lcl}
      {(\cdot)}_{\mathrm{t}} & = & - {(\cdot)}_1 \cos{(2 \alpha)}  - {(\cdot)}_2 \sin{(2 \alpha)} \\
      {(\cdot)}_{\times} & = & + {(\cdot)}_1 \sin{(2 \alpha)} - {(\cdot)}_2 \cos{(2 \alpha)},
\end{array}
\end{equation}
where ${(\cdot)}$ denotes any of $g$, $\gamma$ and $\epsilon$, and $\alpha$ is the azimuthal angle with respect to a chosen center. 

While the cross shear term vanishes for any azimuthally symmetric mass distribution, the
tangential shear can be expressed in terms of the surface mass density as \citep[e.g.][]{1995ApJ...449..460K,2000ApJ...534...34W}
\begin{equation}
\label{eq:gammafrommass}
    \gamma_{\mathrm{t}}(r) = \frac{\overline{\Sigma}(<r)-\overline{\Sigma}(r)}{\Sigma_{\rm{crit}}},
\end{equation}
where $\overline{\Sigma}(<r)$ is the mean surface mass density inside the projected radius $r$, and $\overline{\Sigma}(r)$ is the surface mass density at radius $r$.

\subsection{Simulations}
\label{sect:meth:simu}

We make use of the box2b-hr simulation box from the Magneticum Pathfinder\footnote{\url{http://www.magneticum.org/}} suit of simulations \citep{2014MNRAS.442.2304H,2015ApJ...812...29T,2016MNRAS.463.1797D}, in particular the redshift slice at $z=0.67$ (snapshot 22). The simulation, implemented with magneto-hydrodynamics in the cosmological Smoothed Particle Hydrodynamics (SPH) code GADGET \citep{2001ApJ...549..681S,2005MNRAS.364.1105S}, has a volume of $640^3 h^{-3}$ Mpc$^3$ and $2 \times 2880^3$ particles, and accounts for
cooling, star formation and winds \citep{2003MNRAS.339..312S,2009MNRAS.399..574W};
metals, stellar populations and chemical enrichment \citep{2007MNRAS.382.1050T};
black holes and AGN feedback \citep{2005MNRAS.361..776S, 2010MNRAS.401.1670F};
thermal conduction \citep{2004ApJ...606L..97D};
turbulence \citep{2005MNRAS.364..753D} 
and
passive magnetic fields \citep{2009MNRAS.398.1678D}. The data products were obtained using the Magneticum web portal\footnote{\url{https://c2papcosmosim.srv.lrz.de/}}, using the tools ClusterFind \citep{2017A&C....20...52R} and SMAC \citep{2005MNRAS.363...29D}.

Selecting the 275 most massive halos from the simulation box, the resulting mean and median $\mfh$ are $1.69 \times 10^{14} h^{-1} \msun$ and $1.45 \times 10^{14} h^{-1} \msun$, respectively. After projecting each target onto three mutually orthogonal planes, we extract SZE Compton-Y as well as projected mass images, taking into account simulation particles within $\pm$30 $h^{-1}$ Mpc along the line of sight from the position of the most bound particle. 

We derive the weak lensing shear and convergence, described in \sect~\ref{sect:meth:wl}, from the projected mass at each pixel in the image, and set a constant lensing efficiency $\beta$. The latter has no direct bearing upon the results, as it only affects the noise level in derived masses. 

To emulate the typical resolution of ground-based SZE observations, we convolve the SZE images with a two-dimensional Gaussian of one arcminute full width at half maximum. The peak of a thus convolved image is taken as the SZE center, $\szcentr$. 

\subsection{Miscentering}
\label{sect:meth:misc}

\begin{figure*}
\centering
\includegraphics[width=0.79\columnwidth,clip=True,trim={2 5 0 0}]{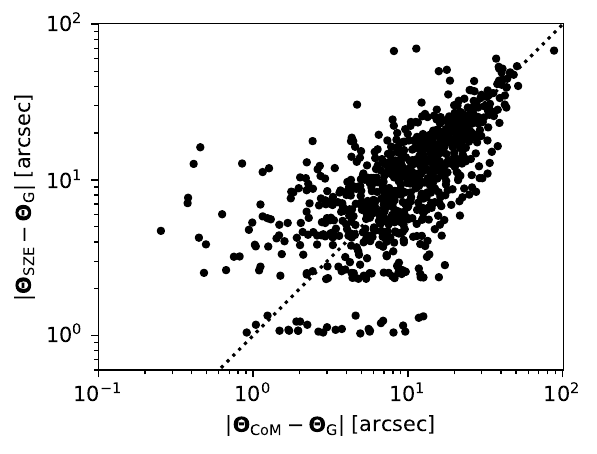} 
\includegraphics[width=0.74\columnwidth,clip=True,trim={2 5 0 0}]{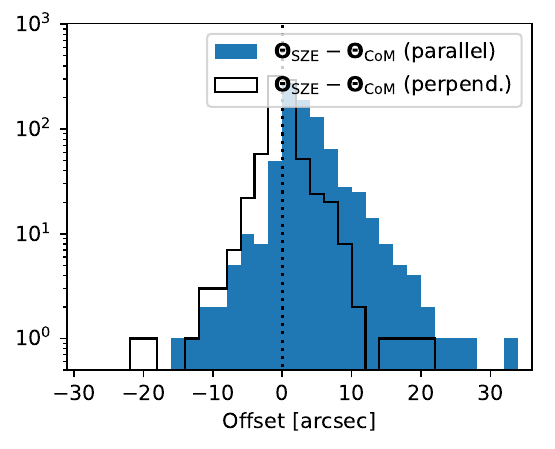}
\includegraphics[width=0.44\columnwidth,clip=True,trim={2 5 5 0}]{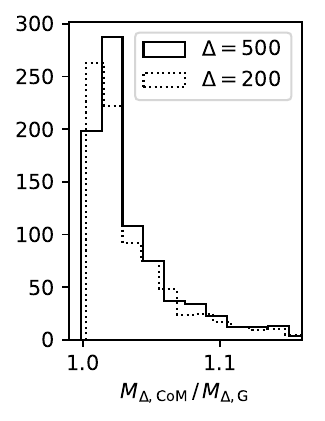}
\caption{\label{fig:szscatter} Left:  projected center offsets of SZE peaks vs. centers of mass, both with respect to the gravitational center, for the 825 targets in the sample. The dotted line indicates the one-to-one-relation. Middle: histogram of the projected offsets between $\szcentr$ and $\mcentr$, decomposed into components respectively parallel and perpendicular to the vector $\mcentr - \gcentr$. Right: Distribution of mass ratios.} 
\end{figure*}

Starting from the position of the most bound particle, which we shall refer to as the gravitational center, or $\gcentr$, we derive the position of the center of mass, $\mcentr$, iteratively. The first estimate is the mass-weighted average position of all particles within a radius $\rfh$ from $\gcentr$. At each new position, we re-compute $\mfh$ and $\rfh$ from the local three-dimensional information on particle positions and masses using the SIMCUT\footnote{\url{https://c2papcosmosim.srv.lrz.de/map/simcut}} tool \citep{2017A&C....20...52R}. Convergence (the projected difference between positions being less than one second of arc) is typically achieved within three iterations. Due to the re-calibration of the center, we must distinguish the mass defined with respect to $\mcentr$, $\mfhm$, from the default mass $\mfhg$, defined in relation to the position $\gcentr$. The difference is small, as indicated in \fig~\ref{fig:szscatter}.

The SZE center $\szcentr$, which we shall use as an observational proxy, is the peak of the convolved SZE image as described in \sect~\ref{sect:meth:simu}. 
\citetalias{2024MNRAS.532.3359S} devised an artificial broadening of the SZE miscentering distribution in order to approximately account for noise components due to primary anisotropies of the cosmic microwave background, millimeter emission from dusty galaxies, and noise from the instrument and Earth's atmosphere. With respect to the issue of isotropic miscentering distributions, however, the latter work reports no significant difference using this broadened miscentering distribution. Thus, we work with noiseless SZE images here.

\begin{figure}
\centering
\includegraphics[width=0.96\columnwidth,clip=True,trim={0 5 0 10}]{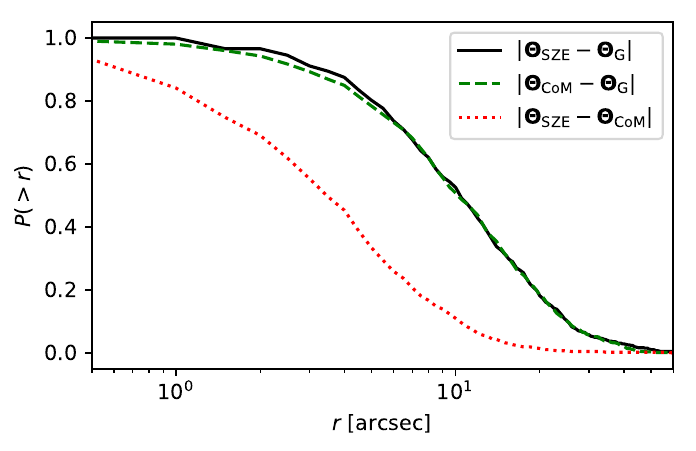} \\
\caption{\label{fig:miscdistros} Azimuthally averaged miscentering distributions, shown as the estimated probability $P(>r)$ that the miscentering amplitude of a given target is greater than a radius $r$. The black solid line and the green dashed line show the miscentering of the SZE peak and the center of mass, respectively, from the gravitational center. The miscentering distribution of the SZE peak with respect to the center of mass is indicated by the red dotted line.} 
\end{figure}

From the projected offsets between the centers defined above, we derive three azimuthally averaged empirical miscentering distributions, namely the distribution describing (i) the SZE peak position $\szcentr$ with respect to the gravitational center $\gcentr$; (ii) the center of mass $\mcentr$ with respect to $\gcentr$; and finally (iii) $\szcentr$ with respect to $\mcentr$. 
These distributions of projected miscentering offsets are shown in angular scale in \fig\ref{fig:miscdistros}. The miscentering distribution $|\szcentr - \mcentr|$ is considerably narrower than the two with $\gcentr$ as reference (the latter two appearing very similar), suggesting a strong spatial correlation between the SZE peak and the center of mass.

In \fig\ref{fig:szscatter}, we show the scatter between the SZE and center of mass miscenterings, indicating that the two are indeed strongly correlated. A different way to view the correlation is to separate the offsets between the SZE peak and the center of mass into orthogonal components, parallel and perpendicular to the direction of the center of mass (with respect to $\gcentr$). 

\begin{figure}
\centering
\includegraphics[width=0.85\columnwidth,clip=True,trim={15 10 35 27}]{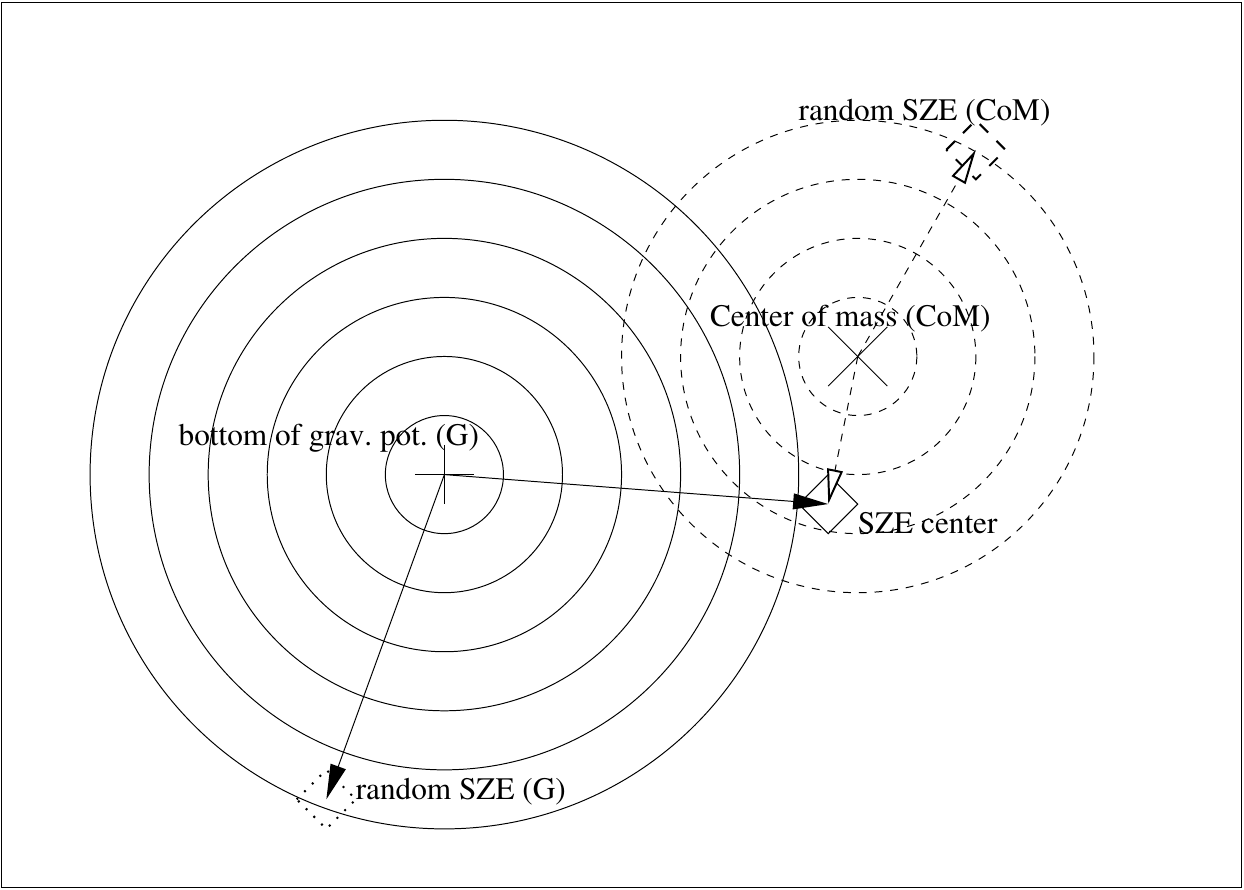} \\
\caption{\label{fig:geom} Schematic view of the miscentering distributions used in this work. Apart from the position of the SZE center taken as the peak SZE emission the simulations, random SZE centers are generated in two ways, namely from the distributions of $|\szcentr - \gcentr|$ and $|\szcentr - \mcentr|$.} 
\end{figure}

In order to test whether miscentering distributions are isotropic, in the sense that the miscentering direction does not systematically alter the estimated weak lensing mass bias distribution, we randomize the distributions $|\szcentr - \gcentr|$, $|\szcentr - \mcentr|$ and $|\mcentr - \gcentr|$
by drawing (with replacement) for each halo a random absolute miscentering from the respective distribution, coupling it with a random angle $\varphi \in [0,2\pi)$ in the sky plane.
The process is schematically illustrated in \fig\ref{fig:geom}. For clarity, we denote the thus randomized miscentering with a subscript (r). 

\subsection{Mass fitting}
\label{sect:meth:mass}

To estimate masses, we use the azimuthally symmetric Navarro-Frenk-White (NFW) density model \citep{1997ApJ...490..493N}, which has shown reasonable consistency with galaxy clusters from both hydrodynamic \citep{2016MNRAS.456.3542T} and n-body \citep{2001MNRAS.321..559B,2012MNRAS.423.3018P,2014ApJ...797...34M,2016MNRAS.457.4340K,2017MNRAS.469.3069G} simulations. In the NFW model, the mass density $\rho(r)$ at physical radius $R$ is given by 
\begin{equation}
    \rho(R) = \frac{M_\Delta}{4\pi f(c_\Delta)} \frac{1}{R(R+\frac{r_{\Delta}}{c_{\Delta}})^2},
\end{equation}
where $c_{\Delta}$ is the so-called concentration parameter, and $f(c_{\Delta}) \equiv \log(1+c_\Delta) - c_\Delta/(1+c_\Delta)$. Projecting the density onto the sky plane yields the mass surface density
\begin{equation}
\label{eq:projectednfw}
    \Sigma(r) = 2 \int_{0}^{\infty} \rho \left( \sqrt{r^2+\zeta^2}\right) \text{d} \zeta,
\end{equation}
where $r$ is the projected radius from the chosen center, and $\zeta$ is in the direction of the line of sight. For the mass surface density and the resulting weak lensing shear, we make use of the analytical expressions derived by \citet{1996A&A...313..697B}.

We tie the concentration parameter $c_{\Delta}$ to 
the mass using the concentration-mass relation of \cite{2015ApJ...799..108D}, with the corrected parameters set of \cite{2019ApJ...871..168D}. While alternative choices of $c_{\Delta}$ strongly affect the normalization of the weak lensing mass bias distribution (e.g. \citealt{2022MNRAS.509.1127S}), we do not expect a significant impact on the \textit{difference} between two distributions derived using different centers.

We fit for the mass of each target using the projected density of \eqn\eqref{eq:projectednfw}
to predict the reduced tangential shear as a function of $\mfh$, radially binning the reduced shear in the simulated images around each chosen center coordinate. In part due to the NFW model being somewhat unsuccessful in predicting the density near the halo center (e.g. \citealt{2018ApJ...859...55C}), and in part to mitigate the effects of miscentering and magnification systematics, shear measurements with small radii are typically excised in weak lensing analyses \citep[e.g.][]{2017MNRAS.465.3361H}. Here we fit to reduced shears with radii $r \in [\rmin,\rmax$] from the center, with the inner radius $\rmin=0.5$ Mpc and the outer radius $\rmax = 1.7$ Mpc (corresponding to 71$^{\prime\prime}$ and 240$^{\prime\prime}$, respectively). The latter value is chosen so as to minimize both the two-halo term (e.g. \citealt{2000MNRAS.318..203S}, \citealt{2005MNRAS.362.1451M}) and errors in the reduced shear --- as shear is a non-local measure, any localized shear image will have large errors towards the outskirts of the image.

As the determination of the weak lensing mass bias has been found not to be sensitive to the level of statistical noise \citep{2022MNRAS.509.1127S}, no noise is added to the reduced shear. Instead, the data are weighted to reflect the growth of radial bins with projected radius.

The weak lensing mass bias $\bias$ is defined as the ratio of measured to reference ("true") mass, where the latter is either $\mfhg$ or $\mfhm$ (\sect~\ref{sect:meth:misc}). The choice of reference mass is not contingent upon the choice of reference center; we are free to define the center coordinate and the reference mass independently of one another.

For each centering convention applied to our sample, we model the distribution of $\bias$ with normal and log-normal distributions. While neither perfectly models the mass bias, \citetalias{2024MNRAS.532.3359S} found that either one can provide a reasonable approximation. 
We use the symbols ($\munormal$,~$\sigmanormal$) to respectively mean the estimated mean and standard deviation in linear space, and similarly ($\mulognormal$,~$\sigmalognormal$) for the log-normal case. 

Given an estimate of the mass bias distribution, a measured mass can be corrected, dividing by random samples from the former and propagating the uncertainties of the measurements. To quantify the difference between a mass bias distribution derived from actual SZE centers and its randomized counterpart, we define $\overcorrection = \langle (M_{\rm{biased}} - M_{\rm{unbiased}}) / M_{\rm{unbiased}} \rangle $ as the expectancy value of the relative over-correction induced by using random miscentering (a negative value of $\overcorrection$ thus corresponds to an under-correction). Here, the subscript 'unbiased' refers only to the anisotropic miscentering. 

\section{Results}
\label{sect:resu}

\begin{table*}
\centering
\begin{tabular}{c l l l r@{$\pm$}l r@{$\pm$}l  r@{$\empt$}l r@{$\pm$}l r@{$\pm$}l r@{$\empt$}l} 
 \hline
& \multicolumn{2}{c}{reference}   &  & \multicolumn{6}{c}{log-normal distribution} &  \multicolumn{6}{c}{normal distribution} \\
\cmidrule(lr){2-3} \cmidrule(lr){5-10} \cmidrule(lr){11-16}
& center & mass & centering  & \multicolumn{2}{c}{$\munormal$} & \multicolumn{2}{c}{$\sigmanormal$}  & \multicolumn{2}{c}{$\overcorrection$} & \multicolumn{2}{c}{$\mulognormal$} & \multicolumn{2}{c}{$\sigmalognormal$}  & \multicolumn{2}{c}{$\overcorrection$} \\
\hline
0 & $\gcentr$ & $\mfhg$ &         $\gcentr$ & $-$0.024&0.006 & 0.167&0.004 &       &      &  0.991&0.006 & 0.179&0.004 &       &      \\ 
\hline
1 & $\gcentr$ & $\mfhg$ &    $\mcentr$ &  0.015&0.006 & 0.167&0.004 &       &      &  1.030&0.007 & 0.191&0.005 &       &      \\ 
2 & $\gcentr$ & $\mfhg$ &   $\miscgmr$ & $-$0.051&0.007 & 0.189&0.005 &  0.068$\empt$&$\pm$0.009 &  0.967&0.007 & 0.187&0.005 &  0.065$\empt$&$\pm$0.010 \\ 
\hline
3 & $\gcentr$ & $\mfhg$ &    $\szcentr$ &  0.005&0.006 & 0.165&0.004 &       &      &  1.020&0.006 & 0.184&0.005 &       &      \\ 
4 & $\gcentr$ & $\mfhg$ &   $\miscgsr$ & $-$0.056&0.007 & 0.187&0.005 &  0.063$\empt$&$\pm$0.009 &  0.962&0.006 & 0.183&0.005 &  0.060$\empt$&$\pm$0.009 \\
\hline
5 & $\mcentr$ & $\mfhg$ &    $\szcentr$ & 0.005&0.006 & 0.165&0.004 &       &      &  1.020&0.006 & 0.184&0.005 &       &      \\ 
6 & $\mcentr$ & $\mfhg$ &   $\miscmsr$ & 0.007&0.006 & 0.169&0.004 & $-$0.002$\empt$&$\pm$0.008 &  1.023&0.007 & 0.190&0.005 & $-$0.003$\empt$&$\pm$0.009 \\  
\hline
7 & $\mcentr$ & $\mfhm$ &    $\szcentr$ & $-$0.034&0.006 & 0.164&0.004 &       &      &  0.981&0.006 & 0.174&0.004 &       &      \\ 
8 & $\mcentr$ & $\mfhm$ &   $\miscmsr$ & $-$0.032&0.006 & 0.167&0.004 & $-$0.002$\empt$&$\pm$0.008 &  0.983&0.006 & 0.178&0.004 & $-$0.003$\empt$&$\pm$0.009 \\ 
\hline
\end{tabular}
\caption{\label{tab:biasresults} Results of fitting log-normal and normal distributions to the mass bias $\bias$, given the reference center, reference mass and type of centering used. Rows are numbered $0-8$. Pairs of rows separated by horizontal lines represent actual vs. randomized (odd and even row numbers, respectively) centerings. Rows 3 and 5 have identical mass bias distributions, as both are derived from actual SZE centers and have the same mass reference. In each even row (excepting row 0), the overcorrection $\tau$, representing the mass overcorrection due to randomized miscentering (with respect to the row immediately above), is reported. }
\end{table*}

\begin{figure}
\centering
\includegraphics[width=0.9\columnwidth,clip=True,trim={0 0 0 0}]{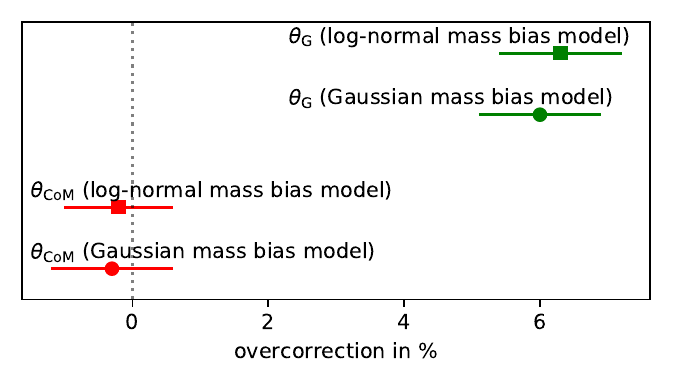} \\
\caption{\label{fig:overcorr} Overcorrection in percent, using different combinations of reference center and mass bias model.} 
\end{figure}

We put the halos from the simulation through the mass fit with four different centering schemes for the reduced shear profile: (1) as a general reference, the gravitational center $\gcentr$; (2) the SZE centers $\szcentr$ as determined from the peak SZE signal; (3) SZE centers randomized from the empirical distribution of $|\szcentr - \gcentr|$; and (4) SZE centers randomized from the empirical distribution of $|\szcentr - \mcentr|$. The resulting estimates for the corresponding weak lensing mass distributions are summarized in \tabl~\ref{tab:biasresults}. A graphical representation of the results is shown in \fig~\ref{fig:overcorr}.

With no miscentering with respect to the gravitational center, masses are underestimated by around 1\% (row zero of \tabl~\ref{tab:biasresults}). Taking the actual SZE centers results instead in an overestimation of 2\% (row 3). We stress that these results are specific to the mass range, redshift and radial weights (in particular, the choice of $\rmin$, $\rmax$ and the assumption of uniformly distributed background galaxies), as well as the choice of model (here, the NFW density profile in conjunction with the chosen concentration--mass relation).  

With $\gcentr$ as the reference, masses corrected by a mass bias distribution $\bias$, derived from the isotropic miscentering $|\szcentr - \gcentr|$, results in a mass overcorrection of $\sim$6\%, independently of which distribution is taken as a model (rows 3 and 4 of \tabl~\ref{tab:biasresults}). This results is fully consistent with \citetalias{2024MNRAS.532.3359S}. When we instead use the center of mass $\mcentr$ as the reference center (rows 5--8), we find a mass overcorrection consistent with zero, with an uncertainty of around one percent. The statement is true whether the reference mass is $\mfhg$ (rows 5--6) or $\mfhm$ (rows 7--8); this is expected as the ratio of the two masses is a constant for each halo in the sample. 

The aforementioned results suggest that systematic errors in the mass bias, resulting from the use of an isotropic miscentering distribution, are at or below one percent when the center of mass is taken as the "true" halo center.

For comparison, we also report the overcorrection in the case where the shear profiles are centered on $\mcentr$, with $\gcentr$ as reference (rows 1 and 2 in the table). Here, the overcorrection is consistent with that derived taking the SZ centers, implying that most of the anisotropic mass bias originates in the difference between $\gcentr$ and $\mcentr$, and that the center of mass is thus a more suitable proxy than the gravitational center for the peak of the surface mass density.

\section{Discussion}
\label{sect:disc}

Considering the significant correlation between $\szcentr$ and $\mcentr$, it is unsurprising that the overcorrection is reduced with $\mcentr$, rather than $\gcentr$, as the reference. Although the absolute miscentering of the SZE peak with respect to the center of mass is reduced by about $50\%$ on average, the overcorrection, attributed to a non-isotropic miscentering distribution, is reduced from $~6$\% to $~ \pm 1$\%. To arrive at this result, we smoothed the SZE signal to a resolution of one arcminute. To see the effects of modifying this scale, we repeat the analysis with the SZE images smoothed to 0.5 arcminutes. While the result is fully consistent in terms of overcorrection, there is a slight broadening (around 6\%) of the SZE miscentering distribution with respect to the center of mass, suggesting that more resolved SZE images would in fact be sub-optimal.  

\citetalias{2024MNRAS.532.3359S} found very similar overcorrection factors from SZE peaks and X-ray centroids. Considering also that X-rays trace mass density of the intra-cluster gas, while the SZE traces gas pressure, we speculate that the center of mass may also serve as a more unbiased mass reference when X-ray centroids are used as the center proxy, although we have not shown this explicitly. 

Our SZE peak miscentering distributions are artificially narrow, as we have not included any noise in the SZE images. This, however, is of little consequence, as an isotropic broadening of the miscentering distribution would not be expected to cause anisotropic effects in the mass bias (see \citetalias{2024MNRAS.532.3359S} for a more in-depth discussion of this statement). 

To determine whether the concentration-mass relation has a bearing on the results, we repeat the analysis with a constant concentration $c_{200} = 3.0$. Though this has an impact on the level of bias, there is no significant change in the overcorrection $\tau$.

We suggest that a relatively simple solution to the problem of non-isotropic miscentering lies in moving the reference center of a halo to the center of mass. As a practical consequence, one would then need to recalculate miscentering distributions with respect to a suitably defined center of mass. A re-definition of the halo mass function, on the other hand, would not be required. 

As an additional test, we repeat the analysis without re-calibrating $\mfh$ when finding the center of mass. Again, we find results fully consistent with those reported in the previous section. As the mass difference is small ($\mfhm$ is 5\% higher than $\mfhg$ on average; see \fig~\ref{fig:szscatter}), this is unsurprising -- the corresponding mean ratio of $\rfh$ is less than 2\%.

\section{Summary and Conclusions}
\label{sect:conc}

Using s snapshot of the Magneticum simulations of massive halos at redshift $z=0.7$, we derived miscentering distributions of the SZE peak in 1 arcminute convolved noiseless SZE images with respect to both the gravitational center and the center of mass. Randomizing both miscentering distributions to make them isotropic by construction, we compared the results to using the actual SZE peaks for each halo. The effect on the weak lensing mass bias, disregarding all other sources of bias, clearly favor the center of mass as the default center proxy, as there is no detectable difference (at the one percent level) in the mass bias whether the randomized or actual miscentering is used. Conversely, a mass overcorrection of approximately 6\% is found when the gravitational center is taken as the center proxy, and an isotropic miscentering distribution is assumed. As a consequence, we suggest that the center of mass, rather than the bottom of the gravitational potential, represented in simulations by the most bound particle in a halo, may for practical reasons be a more robust marker of the halo center when estimating miscentering distributions from cosmological simulations, while the halo mass function needs not be modified. 
We speculate that a similar approach may be used when X-ray centroids are the center proxy of choice.  

\section*{Acknowledgements}

We would like to thank Klaus Dolag, Antonio Ragagnin, Alex Saro and Veronica Biffi for their help in using and interpreting the Magneticum Pathfinder simulations. We thank Peter Schneider for helpful discussions, and the anonymous referee for useful inputs improving the manuscript.

We acknowledge support from the German Federal Ministry for Economic Affairs and Energy (BMWi) provided through DLR under projects 50OR2002, 50OR2106, 50OR2302,
and 50QE2002 as well as support provided by the Deutsche Forschungsgemeinschaft (DFG, German Research Foundation) under grant 415537506. 

The Innsbruck authors acknowledge support provided by the Austrian Research Promotion Agency (FFG) and the Federal Ministry of the Republic of Austria for Climate Action, Environment, Mobility, Innovation and Technology (BMK) via the Austrian Space Applications Programme with grant numbers 899537, 900565, and 911971.  

%%%%%%%%%%%%%%%%%%%%%%%%%%%%%%%%%%%%%%%%%%%%%%%%%%
\section*{Data Availability}

The data underlying this work will be shared on reasonable request to the corresponding author.

%%%%%%%%%%%%%%%%%%%% REFERENCES %%%%%%%%%%%%%%%%%%

\bibliographystyle{mnras}
\bibliography{com} 

%%%%%%%%%%%%%%%%% APPENDICES %%%%%%%%%%%%%%%%%%%%%

%\appendix

%\section{Some extra material}

%%%%%%%%%%%%%%%%%%%%%%%%%%%%%%%%%%%%%%%%%%%%%%%%%%

% Don't change these lines
\bsp	% typesetting comment
\label{lastpage}
\end{document}